\documentclass[twocolumn,prl,showpacs,preprintnumbers,superscriptaddress,amsmath,amssymb]{revtex4}
\usepackage{amsmath,dcolumn,graphicx,epsf,ulem}         
\setkeys{Gin}{width=8.5cm,keepaspectratio}         
\usepackage{dcolumn}% Align table columns on decimal point         
\usepackage{bm}% bold math         

\begin{document}         
         
\title{Structure and magnetism in $\rm LaCoO_3$}

\author{D.~P.  Belanger}
\affiliation{Department of Physics, University of California, Santa Cruz, CA 95064, USA}   
\author{T. Keiber}
\affiliation{Department of Physics, University of California, Santa Cruz, CA 95064, USA}   
\author{F. Bridges}
\affiliation{Department of Physics, University of California, Santa Cruz, CA 95064, USA}   
\author{A. M. Durand} 
\affiliation{Department of Physics, University of California, Santa Cruz, CA 95064, USA}   
\author{A. Mehta}
\affiliation{Stanford Synchrotron Radiation Lightsource, SLAC National Accelerator Laboratory, Menlo Park, California 94025, United States}   
\author{H. Zheng}
\affiliation{Materials Science Division, Argonne National Laboratory, Argonne, IL 60439, USA}   
\author{J.~F. Mitchell}
\affiliation{Materials Science Division, Argonne National Laboratory, Argonne, IL 60439, USA}   
\author{V. Borzenets}
\affiliation{Stanford Synchrotron Radiation Lightsource, SLAC National Accelerator Laboratory, Menlo Park, California 94025, United States}   
   
\date{\today}   
   
\begin{abstract}   
   
The temperature dependence
of the hexagonal lattice parameter $c$
of single crystal $\rm LaCoO_3$ (LCO)
with $H=0$ and $800$~Oe, as well as LCO bulk powders
with $H=0$, was measured using high-resolution
x-ray scattering near the transition temperature
$T_o\approx 35$~K.  The change of $c(T)$
is well characterized by a power law in
$T-T_o$ for $T>T_o$ and by a temperature
independent constant for $T<T_o$ when convoluted with
a Gaussian function of width $8.5$~K.
This behavior is discussed in the context of the
unusual magnetic behavior observed in LCO as well
as recent generalized gradient approximation calculations.
   
\end{abstract}   
   
\maketitle         
        
$\rm LaCoO_3$ (LCO) exhibits
magnetic behaviors
that cannot be explained using
local Co ion spin state transitions alone.
Experiments and calculations have shown
the importance of extended states~\cite{sjabbbmpz09,jbsbamz09,mlzmfhb12,lh13}
and rhombohedral distortion near a transition
at $T_o \approx 35$~K that involves the
Co-O-Co bond angle.
A core-interface model was developed~\cite{dhbcyfab15,dbhycfbab15}
that includes strained interface regions
near surfaces and interfaces
with impurity phases, particularly $\rm Co_3O_4$,
and core regions far from such surfaces
and interfaces.  In this model, ferromagnetic long-range
order below $T_C=89$~K is associated with the interface regions.
This is similar to the case of thin films which have strained interfaces
with substrates that generate ferromagnetic moments.~\cite{mbjavs15}
For bulk particles and single crystals, most of the spins are
in the core regions that are dominated by antiferromagnetic
interactions, though the core regions never develop antiferromagnetic
or ferromagnetic long-range order.

Neither indirect~\cite{psmmpcthk06}
nor direct~\cite{jbsbamz09,sjabbbmpz09} experimental observations yielded any evidence of
a Jahn-Teller distortion of the oxygen octahedra surrounding Co ions.
This eliminated one model proposed~\cite{kesaks96} that would have accounted
for the magnetic behavior or LCO through modification of the local spin
excitations, allowing excitations of order 100~K in local-density
approximation calculations to explain the magnetic behavior near $100$~K.
On the other hand, generalized gradient approximation (GGA) calculations~\cite{lh13}
show that extended states can account for excitations in the required energy range
without the Jahn-Teller distortion of the oxygen octahedra and this is
in better agreement with the experimental observations.  As the temperature decreases from room
temperature, the behavior of the inverse magnetization, $H/M$ vs $T$,
indicates that antiferromagnetic correlations
in the core region increase below $T_C$, reach a broad
maximum near $T=40$~K and then decrease substantially.~\cite{dhbcyfab15}

It was shown in neutron scattering studies~\cite{dhbcyfab15,dbbycfb13} that,
as the temperature decreases towards $T_o$,
the Co-O-Co angle $\gamma$ rapidly
approaches an angle $\gamma = 163^{\circ}$.
Below $T \approx 40$~K, little change in the angle was observed.
Generalized gradient approximation (GGA) calculations ~\cite{lh13}
indicate LCO to be non-magnetic for
$\delta y = \frac{d}{a}\cos(\gamma/2)$ above
a critical value of $0.052$,
where $a$ is the hexagonal lattice parameter and $d$ is
the Co-O bond length. This critical value
for $\delta y$ corresponds
to $\gamma _C = 163^{\circ}$ using the measured
low $T$ bulk powder values $a=5.40$~\AA\ and $d=1.915$~\AA.
~\cite{dbbycfb13} This agrees with the
experimental observations~\cite{dbbycfb13} for $T \le 40$~K.

To further explore the relationship between
the structural changes and the magnetic
behavior, we have used high resolution
x-ray scattering techniques.
Although x-ray scattering is less sensitive to the oxygen positions
and is usually less reliable to directly measure the Co-O-Co bond angle,
it was observed~\cite{dbbycfb13} that the transition near To is also expressed
in the behavior of all the lattice parameters.  Hence,
the transition can be studied with high resolution using,
for example, x-ray measurements of the temperature dependence
of the $c$ hexagonal lattice parameter.
We have used high resolution x-ray scattering measurements
of $c$ versus $T$ to better characterize the nature
of the transition at $T_o$.  In addition, we have
explored the dependence of the transition upon application
of a small magnetic field in single crystals and we have compared the behavior
of single crystals and a bulk LCO powder. 
From the results of the experimental measurements, we propose
a model to understand how LCO develops from the low $T$
non-magnetic state to a magnetic state for $T>40$~K.

The bulk particles of the powder samples were synthesized using a standard solid state reaction.
Bulk $\rm Co_3O_4$ and $\rm La_2O_3$ were ground together using a mortar and
pestle and then heated to 850$^\circ$C - 1050$^\circ$C for approximately
8 hours in air. The grinding and heating processes were repeated five times,
with a final 24 hour heating at 1100$^\circ$C.  Neutron
powder diffraction characterizations indicate the primary
phase is $\rm LaCoO_3$, with a small amount, $4.5(4)$\% by weight,
of the impurity phase $\rm Co_3O_4$.

To synthesize the two single crystals, labeled A and B, a
stoichiometric powder of $\rm LaCoO_3$ was prepared from
$\rm La_2O_3$ (prefired at 1000$^\circ$C immediately prior to use)
and $\rm Co_3O_4$, pressed into rods, and
sintered in air.  Single crystals of $\rm LaCoO_3$ were then grown from the
rods using an IR optical-image floating zone technique (model NEC
SC15-HD)  in a 100\% $\rm O_2/Ar$ atmosphere.  Samples were found to be single
phase $\rm LaCoO_3$ by powder x-ray diffraction of a pulverized section of
the crystal. 

\begin{figure}
\includegraphics[width=2.3in, angle=270]{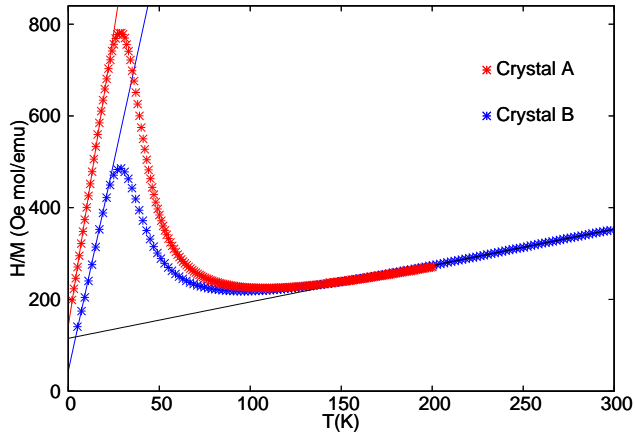}
\caption
{$M/H$ vs $T$ for two single crystals, A and B (see text), with $H=1$~T along with
Curie-Weiss fits (Eq.~\ref{eq:CW}) for $170<T<300$~K and for low temperature.
\label{fig:HM}
}
\end{figure}

\begin{figure}
\includegraphics[width=2.3in, angle=270]{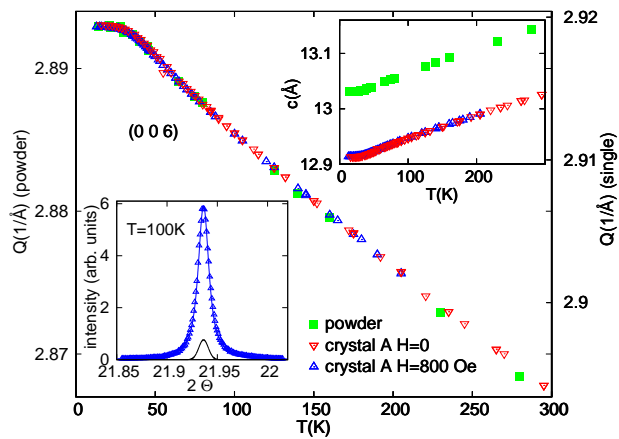}
\caption
{The temperature dependence of Q for crystal A,
for $H=0$ and $H=800$~Oe,
and the bulk powder for $H=0$, determined from the
(0 0 6) Bragg positions. The upper insert shows the
 hexagonal $c$ lattice parameters.  There is no
 observed difference between cooling and heating data.
 The lower insert shows a fit to the very sharp (0 0 6) diffraction peak
 using a sum of a Lorentzian and a Gaussian; the small Gaussian contribution
 is below the data and combined fit.
 \label{fig:006}
 }
\end{figure}

\begin{figure}
\includegraphics[width=2.2in, angle=270]{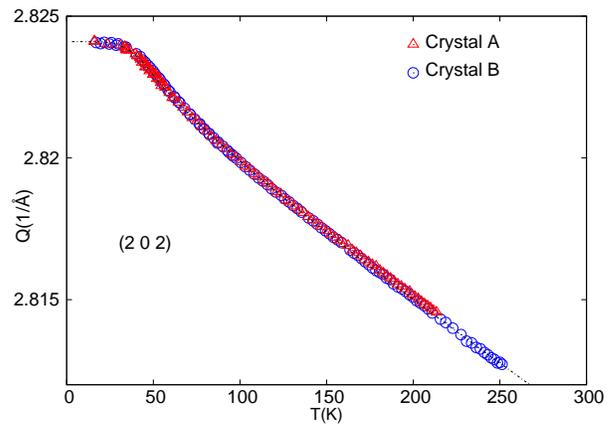}
\caption
{The temperature dependence of Q for
crystals A and B, determined from the
(2 0 2) Bragg positions.  There is no
observed difference between cooling and heating data.
}
\label{fig:202}
\end{figure}

The magnetization of crystals A and B was measured using a Quantum Design 9T Physical Properties
Measurement System (PPMS).  No attempt was made to align the crystals
for the measurements, as they are heavily twinned.
The inverse magnetization, shown for $H=10$~kOe as $H/M$ vs $T$ in Fig.~\ref{fig:HM}, indicates,
from the straight line behavior for $170<T<300$,
that the single crystals are dominated by antiferromagnetic interactions, consistent with
bulk powders measured previously.  A fit of the Curie-Weiss behavior,

\begin{equation}
M/H=\frac{C}{T-\theta_{CW}},
\label{eq:CW}
\end{equation}

\noindent {to the data from crystal B for $170<T<300$~K yields
$C=1.26 \pm 0.02$~emu$\cdot$K/mol and $\theta_{CW} = -145 \pm 5$~K,
which suggests a slightly weaker antiferromagnetic system than the
the bulk powder, which gave $C=1.49 \pm 0.02$ emu$\cdot$K/mol and $\theta_{CW} = - 182 \pm 4$~K
in a previous bulk powder study.~\cite{dbbycfb13}
The average effective moment is $\mu_{eff} = 3.17(2)$ $\mu_{B}$
per Co ion, using $\mu_{eff}^2/\mu_{B}^2=3k_{B}C/N_A$,
where $\mu_B$ is the Bohr magneton, $N_A$ is Avogadro's
number, and $k_B$ is the Boltzmann constant.  This is smaller than the
value $\mu_{eff} = 3.45(2)$ $\mu_{B}$ obtained for a LCO bulk powder.~\cite{dbbycfb13}
The negative value of $\theta_{CW}$ indicates antiferromagnetic interactions and,
from its magnitude, the system might be expected to exhibit antiferromagnetic
long-range order below $T \approx 100$~K, but that is never observed.
Below $T \approx 30$~K, a fit to the Curie-Weiss
expression yields $\theta_{CW} = - 5.5 \pm 0.5$~K
and $C=0.39 \pm 0.02$~emu$\cdot$K/mol for Crystal A and $\theta_{CW} = - 2.4 \pm 0.5 $~K
and $C=0.55 \pm 0.02$~emu$\cdot$K/mol for Crystal B.  The various
Curie-Weiss results are summarized in Table~\ref{Table:CW}.  Unlike the parameters obtained for $T>170$~K,
where the interactions are sufficiently weak,
the Curie-Weiss parameters at low $T$ are not directly interpretable in terms of the moment and 
interaction strength because the short-range correlations are not weak over the range $10<T<35$~K.
Furthermore, the behavior is field-dependent.~\cite{dhbcyfab15}
Rather, this behavior likely represents a decrease in magnetic long-range correlations
and a rapid approach to the $T=0$ susceptibility.
The low $T$ state is more complex than would be predicted in a $S=0$ local Co spin state
model and low temperature models have been proposed.~\cite{epbzml15,psmmpcthk06}
It was shown~\cite{dhbcyfab15} for bulk particles that the height of the peak in $H/M$ vs $T$
increases with a decrease in the dominant impurity phase $\rm Co_3O_4$ and, while $\rm Co_3O_4$ is likely absent in the
single crystals, the difference in peak height between crystal A and B is probably a result of the presence of fewer
twinning planes and other defects in crystal A.  Consistent with the interpretation that crystal A is higher in quality,
crystal A exhibits fewer strong scattering peaks from crystal twinning relative to crystal B.

\begin{table}
\caption{Curie-Weiss parameters using Eq.~\ref{eq:CW} for crystals A and B and the bulk powder, measured
with $H=1$~T.
\label{Table:CW}
}

\begin{tabular}{c | l*{4}{c}}
	& & \textbf{A} & \textbf{B} & \textbf{bulk powder} \\
\hline
& & & {$\mathbf{170<T<300}$~\textbf{K}} \\%[0.15cm]
\hline
$C$~(emu$\cdot$K/mol) & & - & 1.26(2) & 1.49(2) \\
$\theta_{CW}$~(K) & & - & -145(5) & -182(4) \\
$\mu_{eff}$~($\mu_B$) & &-& 3.17(2)  & 3.45(2)  \\[0.2cm]
\hline
& & & {$\mathbf{T<30}$~\textbf{K}} \\%[0.15cm]
\hline
$C$~(emu$\cdot$K/mol) & & 0.39(2) & 0.55(2) & - \\
$\theta_{CW}$~(K) & & -5.5(5) & -2.4(5) & -\\
\hline
\end{tabular}
\end{table}

\begin{figure}
\includegraphics[width=3.2in, angle=0]{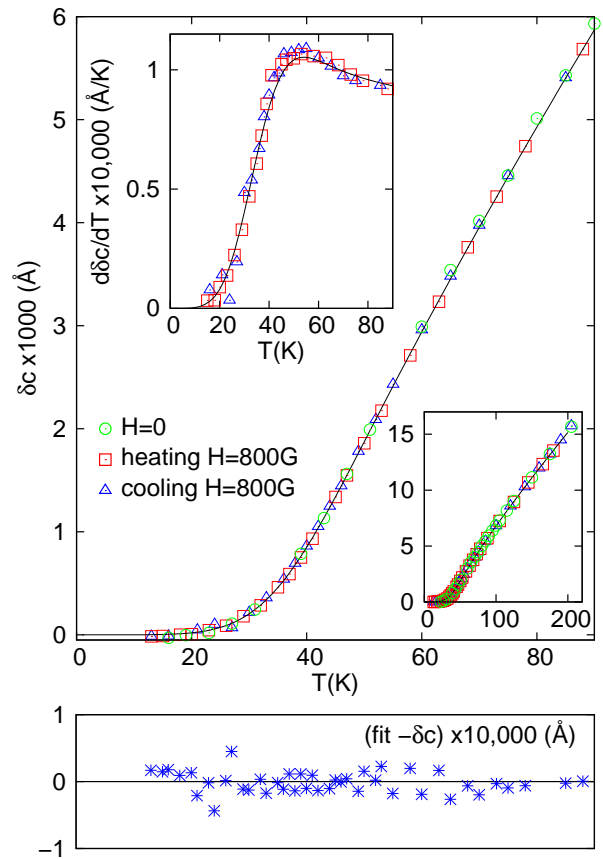}
\caption
{The change in the $c$ lattice parameter with
temperature with a fit to Eq.~\ref{eq:power} shown by
the solid curve, for crystal A with $H=0$ (circle) and $800$~G (squares/triangles).
There is no significant difference between
the cooling and heating data.  The upper
insert shows the derivative of the change in
$c$ vs $T$ (same point symbols) and the derivative of the fit shown
in the main figure for $H=800$~G.  The lower insert is the
same as the main figure, but extending to higher
temperature.  The bottom panel shows the fit
minus the data (times $10^{4}$) for the $H=800$~G data shown in the main
figure.
}
	\label{fig:trevor}
\end{figure}

\begin{figure}
\includegraphics[width=3.5in, angle=0]{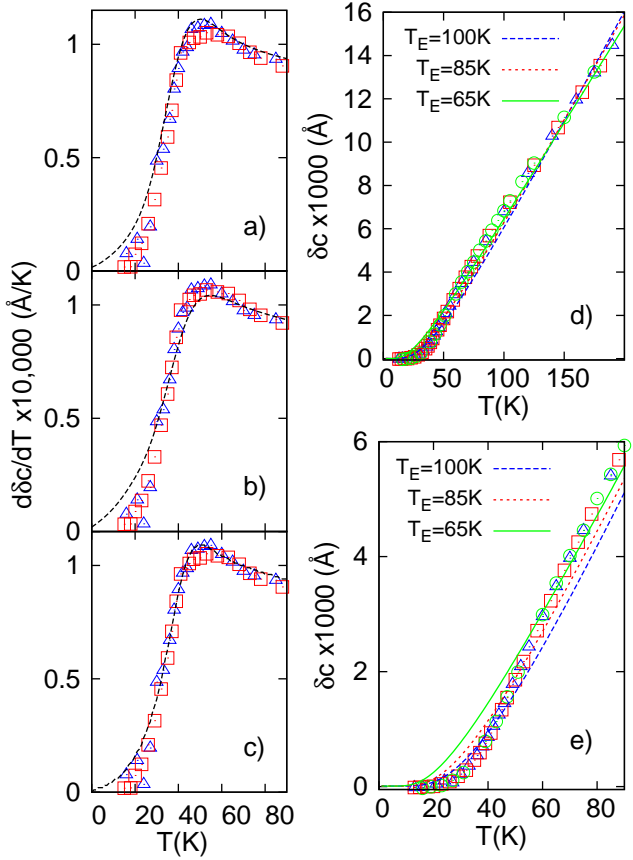}
\caption
{Fits to the derivative data shown in the upper insert of
Fig.~\ref{fig:trevor} using
a) a power law above and below $T_o$ with a negative
exponent, b) a power law above and below
$T_o$ with a postive exponent, and c) an exponential function below
and a power law above $T_o$, as well as
fits of the data shown in the main figure
to an Einstein function in d) and e).
The details of the fits are discussed
in the main text.
}
	\label{fig:thermal}
\end{figure}

High resolution x-ray data were collected as a function of temperature on
beam line 10-2 at SSRL using either 15 or 20 keV photons. The twinned
single crystal samples were mounted on a thick copper block, with the (111) axis
of the pseudo-cubic cell vertical; all reflections were in a vertical plane.
For twinned crystals the (006) hexagonal ($c$ axis) reflection and the (202) from another
twin are close in angle, but were easily resolved. A Sunpower Stirling cryocooler (CryTel-MT)
%--- [need to add this detail]
cryostat was used with a calibrated thermometer
mounted on the copper block close to the sample. Each time the temperature was
changed, it took 10-12 minutes for equilibrium to be established such that no
hysteresis was observed. At each temperature, the full line shape was recorded,
(e.g.\ inset of Fig.~\ref{fig:006}).
For the measurements in the applied field, permanent magnets were mounted close
to the sample on opposing sides.

We studied the temperature behavior of the $c$ lattice parameter of LCO,
using the $\rm R\bar{3}c$ crystal structure, by examining the
hexagonal lattice (0 0 6) Bragg peak as a function of
temperature for crystal A and the bulk powder, as shown in Fig.~\ref{fig:006}.  For
the crystal, we took measurements in zero field and in a field $H = 800$~Oe.
This field strength is known to be sufficient to suppress much of the ferromagnetic
ordering in bulk powders~\cite{dhbcyfab15}.  However, no difference is discernible between
the data with 800~Oe applied perpendicularly to the $c$ axis and data taken in zero field.
We also observed no difference between cooling and heating.
Powder data are also shown in Fig.~\ref{fig:006}
and have nearly the same temperature dependence as the single crystal, but with a slightly
smaller overall value for $Q$, the magnitude of the scattering vector.
The inset of Fig.~\ref{fig:006} shows the lattice parameter
$c=12\pi/Q$ vs $T$; the $c$ parameter for the powder is larger than that of the single crystal.

For crystals A and B in zero field, $Q$ vs $T$ at
the (2 0 2) Bragg peak is shown in Fig.~\ref{fig:202}.  No significant difference between the two crystals
is discernible and there is no observed difference between
data taken upon cooling and heating.  It is not straightforward to interpret the
shape of the curves for this Bragg point because it involves both
the $a$ and $c$ lattice parameters.

We find that we can obtain an excellent fit to the $H=0$ and $800$~Oe $c$ vs $T$ data from crystal A
using a power law in $T-T_o$ for $T>T_o$ and no
change for $T<T_o$ convoluted with a Gaussian function,

\begin{equation}
\delta c = \frac{b}{\sqrt{2\pi}\sigma}\int_{-\infty}^{\infty} t^{1+\beta}\exp(-x^2/2\sigma^2)dx
\label{eq:power}
\end{equation}

\noindent where
\begin{equation}
t=(T-T_o-x)/T_o \quad ,
\label{eq:reduced}
\end{equation}

\noindent and where $T_o=35(2)$~K, $\sigma = 8.5(1.0)$~K, $b=4.0(2)\times 10^{-4}$~\AA\ for $T \ge T_o$ and $b=0$ for $T \le T_o$, and $\beta=-0.13(1)$.
The Gaussian convolution roughly accounts for rounding of the power law behavior.
The data and fit are shown in Fig.~\ref{fig:trevor} for $0<T<80$~K.  The lower insert of the main
figure shows the same fit over a wider temperature range $0<T<200$~K.
The upper inset shows $d(\delta c)/dT$ vs $T$ for $H=800$~Oe and the temperature derivative of the fit shown in the
main figure.  The bottom panel shows the fit minus the data multiplied by $10^{4}$.
The fits are excellent representations of the observed $T$ dependence of $c$
and, along with recent GGA calculations~\cite{lh13} suggest
the model for the behavior discussed below.

To investigate whether using a power law or exponential temperature
dependence
below $T_o$ would be consistent with the data,
we tried several other fits to the same data and examples are shown
in Fig.~\ref{fig:thermal}. For the comparison, we fit the temperature derivative
of $\delta c$ vs $T$ using

\begin{equation}
	\frac{d}{dT}(\delta c) = \frac{1}{\sqrt{2\pi}\sigma}\int_{-\infty}^{\infty} (at^{\beta}+d)\exp(-x^2/2\sigma^2)dx
	\label{eq:power_2}
\end{equation}

\noindent for $T>T_o$ and, for $T<T_o$,
\begin{equation}
	\frac{d}{dT}(\delta c) = \frac{1}{\sqrt{2\pi}\sigma}\int_{-\infty}^{\infty} (a'(-t)^{\beta}+d')\exp(-x^2/2\sigma^2)dx
	\label{eq:power_3}
\end{equation}

\noindent with the results shown in Table \ref{Table:params} for the cases where $\beta < 0$ and $a'=0$ (column A), $\beta < 0$ and $a'\ne 0$ (column B
and panel a of Fig.~\ref{fig:thermal}), and $\beta > 0$, with $d=d'$
for continuity at $T_o$ and $d'=-a'$ to ensure that $\frac{d}{dT}(\delta c)=0$ at $T=0$ (column C and panel b of Fig.~\ref{fig:thermal}).
We also include (column D and panel c of Fig.~\ref{fig:thermal}) a fit where Eq.~\ref{eq:power_3} was replaced for $T<T_o$ by,
\begin{equation}
	\frac{d}{dT}(\delta c) = \frac{1}{\sqrt{2\pi}\sigma}\int_{-\infty}^{\infty} a'(\exp(-h/T)-1)\exp(-x^2/2\sigma^2)dx
\label{eq:exp_1}
\end{equation}

\noindent where, for continuity at $T_o$, we set $d = a'(\exp(-h/T_o)-1)$. 
This is to explore whether an exponential function might work better for $T<T_o$.  The results are shown in panel c.
The fits shown in panels a, b and c were done over the temperature range $10<T<200$~K, but are
shown, for clarity, over the smaller range $10<T<80$~K.  Well above $T>80$ K, the three fits are excellent and are indistinguishable.
In each of these three fits, the rounding was set to be $\sigma = 8.5$~K, the value
used in the fit shown in Fig.~\ref{fig:trevor}.  No significant improvement was found by varying the
amount of rounding and the rounding could not be reduced significantly without reducing the
quality of the fit.  The second and third columns of fitting results of Table \ref{Table:params} are similar,
but with opposite signs for the exponent $\beta$.  The significant rounding makes it difficult to distinguish
between a positive and negative value for $\beta$; using a positive $\beta$ yields a similar
quality fit (panel a of Fig.~\ref{fig:thermal} to that shown in Fig.~\ref{fig:trevor}.
Panel a and b of Fig.~\ref{fig:thermal}, corresponding to the second and third column of fitting results in Table~\ref{Table:params},
show that adding a low $T$ power law results in a fit of poorer quality at low $T$.
Panel c of Fig.~\ref{fig:thermal}, corresponding to the last column of fitting results in Table~\ref{Table:params},
illustrates the inclusion of an exponential low $T$ behavior.  Although this yields better results than a low $T$ power
law behavior, the fit is clearly less good than the fits done with no function below $T_o$.

\begin{table}
	\caption{Fit parameters using Eq.\ ~\ref{eq:power_2}, \ref{eq:power_3} and \ref{eq:exp_1} with fixed
		$\sigma=8.5$~K, as described in the text.  The units are K for $T_o$ and $h$ and \AA$/$K for $a$, $a'$, $d$ and $d'$.
		\label{Table:params}
	}
	\begin{tabular}{c|l*{6}{c}}
		& \textbf{A} && \textbf{B} && \textbf{C}  && \textbf{D} \\
			& Eq.~4 && Eq.~4 \& 5 && Eq.~4 \& 5 && Eq.~4 \& 6 \\
		\hline
		$T_o$        &  35(2) && 40(2) && 41(2) && 44(2) \\
		$\beta$        &  -0.13(1) && -0.12(1) && 0.26(1) && 0.275(10) \\
		$a (\times 10^5)$         & 9.86(30)   && 9.6(5)  && -4.25(30)  && -1.35(10)  \\
		$a' (\times 10^5)$         & -  && 20(1) && -1.38(10)  && 0.45(3) \\
		$h$         & -  && - && - && 12.9(5) \\
		$d $         & 0  && 0 && $-a'$ && $a'(e^{-h/T_o}-1)$ \\
		$d' $         & -  && $-a'$ && $-a'$ && - \\

	\end{tabular}
\end{table}

The large parameter
$\sigma = 8.5(1.0)$~K in the Gaussian convolution is
a physically meaningful parameter intrinsic to LCO.
It cannot be a description of rounding caused by impurities, defects,
surfaces, or interfaces because
the fits are essentially the same for both single crystals, which vary
significantly in quality, and the bulk powder.

It was previously shown, using neutron scattering results,~\cite{dbbycfb13} that an Einstein-like
function~\cite{aynctshok94} of the form

\begin{equation}      
	y(T) = y(0)[1 + \alpha(\coth(\frac{T_E}{2T}) - 1)]
	\label{eq:Grun}      
\end{equation}

\noindent is not as good a description of the data as the power law behaviors.  We show such fits to the
high resolution x-ray data for $\delta c$ vs $T$ in
panels d and e of Fig.~\ref{fig:thermal} with three different values of $T_E$.  Over the large
temperature range shown in panel d, the fits appear fairly good, but the same fits shown in panel e
over the temperature range $10<T<90$~K demonstrate that, in the
region near $T_o$, the fits are quite poor for any reasonable value of $T_E$.

\section*{Conclusions}

We have demonstrated that the temperature dependence of the
$c$ lattice parameter in LCO single crystals for
$T>T_o$ is well described by a power law behavior in the parameter
$T-T_o$ and essentially no temperature change for $T<T_o$,
when convoluted with a Gaussian
function with a significant width of $8.5$~K.
This behavior is not altered by a field $H=800$~Oe applied
perpendicular to $c$.
The behavior, including the amount of rounding, is nearly the
same in two crystals that exhibit a different degree of
crystallinity.  Bulk LCO particles show the same
behavior in the change of $c$ with temperature,
though the $c$ parameter itself is slightly larger in the particles.
We observed no significant
difference upon cooling and heating in any of the samples.
We conclude that, for large crystals and bulk powders, the transition
at T = To is insensitive to small applied fields, differences in crystallinity,
thermal cycling procedures or, with respect to the
temperature dependence, the size of the system.
On the other hand, previous experiments~\cite{dbhycfbab15} using very small nanoparticles,
where all spins are near
particle surfaces, showed that as the core regions disappear, the transition
near $T_o$ does as well,~\cite{dbhycfbab15} demonstrating that the transition is
associated with the core regions far from
interfaces.

The ferromagnetism
in single crystals~\cite{yzg04_b} and bulk powders~\cite{dhbcyfab15}
is very different, being much smaller in the single crystals and
exhibiting a larger difference between zero-field cooling and
cooling in a field.
The dependence of the ferromagnetic moment on the
abundance of surfaces and impurity interfaces, including twin boundaries,
shows that ferromagnetism is associated~\cite{dhbcyfab15,dbhycfbab15}
with them and not the core regions. Hence, the transition
in the core regions is unaffected
by the ferromagnetic moment, which is why suppressing
the ferromagnetism with the applied field
has no discernible effect on the transition observed
near $T_o$.

Recent results of GGA calculations~\cite{lh13}
give clues to the possible origin of the transition near
$T_o=35$~K.  First, the energy was found to be minimized for
$\delta y \approx 0.052$.  Second, at this value of
$\delta y$, a magnetic state is approximately
$3.2$~meV/Co above a nonmagnetic state.
Although the calculations
indicate a ferromagnetic
moment forming in the magnetic state, we note
that an antiferromagnetic interaction dominates
in the experiments, as indicated
by the Curie-Weiss fits, and, in the core region,
neither antiferromagnetic nor ferromagnetic long
range ordering is observed for $T>10$~K~\cite{dhbcyfab15}.
Magnetic correlations are observed for
$\delta y < 0.052$ in the experiments.
Finally, the magnetic state energy is decreased
as $\delta y $ decreases.  We describe a model
consistent with these results as well as those
of the experiments.

At low temperature, the system
is nonmagnetic and, because the value
$\delta y \approx 0.052$ represents the minimum energy,
it does not change with temperature.
Previous inelastic neutron scattering experiments have been interpreted in terms
of a thermally activated process;~\cite{psmmpcthk06,plrlqcozmcsm06} a fit of the
inelastic neutron peak intensity vs T to an exponential thermal activation
of the form $I \propto \exp(-E/k_BT)$ 
yields an activation temperature of 120~K.~\cite{psmmpcthk06}
The GGA calculations predict an
activation temperature of approximately $37$~K.
We see from the fits of the data shown in Fig.~\ref{fig:thermal}
that an exponential fit tends to yield an activation temperature
significantly higher than the $T_o$ used in
power law fits.  As such, the higher characteristic
temperature from the inelastic scattering fit 
is not necessarily inconsistent with the data and the
power law fits obtained in this study or the GGA calculation 
results.  The similarity of the inelastic scattering
fits and the fits in panels d and e in Fig.~\ref{fig:thermal}
suggest that the inelastic scattering and structural phenomena
are related.
It is clear that the magnetic state is thermally activated,
likely starting at local sites as $T$ increases.
As $T$ continues to increase, more sites are thermally
activated to the magnetic state and they become
significant in number
when $T$ becomes comparable to the activation energy
$3.2$~meV/Co indicated by the calculations.~\cite{lh13}
Note that 3.2 meV corresponds to a
temperature of $37$~K, which is
very close to the experimental value $T_o = 35$~K
for the transition given by the power law
fits shown in Fig.~\ref{fig:trevor} and Table~\ref{Table:params}.
At $T=T_o$, the magnetic
sites are dense enough to percolate, forming
a long range cluster.
A power law behavior in $T-T_o$ is typical
of cluster percolation,~\cite{sa94}
but the transition at the percolation threshold
is clearly rounded.

The activated magnetic regions will
tend to have a larger bond angle.~\cite{lh13}
Because the experiments measure the average lattice parameters,
two effects could lead to the observed
rounding of the transition.  First, the lattice
strain caused by the formation of magnetic clusters
is relatively long-ranged and the local lattice
parameters associated with magnetic and nonmagnetic
are not distinct values but rather vary with
the proximity to the boundaries of the clusters.
Second, the magnetic sites
are not randomly excited, but rather, to reduce the
strain energy, are more likely next to other magnetic sites.
These two effects from the strain energy should result
in a deviation from the power law behaviors characteristic of random
site percolation~\cite{sa94} when $T$ is close to $T_o$.
The large value of the rounding reflects the large
amount of lattice strain from the lattice
mismatch of the nonmagnetic and magnetic regions.

We have shown that the temperature dependence of the
lattice parameter $c$ can be modeled by a nonmagnetic
state at low $T$ and an activated magnetic state
with a percolation threshold for magnetic clusters
near $T_o$.  To better model the percolation of magnetic clusters
in LCO, one would need to consider the competing energies
and bond angles between the nonmagnetic and magnetic
regions, the magnetic ordering energy, the lattice strain energies,
and the dependence of the average lattice parameter $c$
on the inhomogeneous bond angle distribution.

Use of the Stanford Synchrotron Radiation Lightsource, SLAC National
Accelerator Laboratory, is supported by the U.S. Department of Energy, Office
of Science, Office of Basic Energy Sciences under Contract No.
DE-AC02-76SF00515.  Work at Argonne National Laboratory (crystal growth and characterization)
was supported by the U.S. Department of Energy, Office of Science,
Basic Energy Sciences, Materials Science and Engineering Division.

\bibliography{magnetism_thesis}

\end{document}